\documentstyle[preprint,prd,aps,eqsecnum,floats,epsfig]{revtex}
\tighten
\begin{document}
%

%
\title{Delta isobar masses, large $N_c$ relations, and the quark model}
%
\author{Simon~Capstick}
\address{
	Department of Physics
	\& Supercomputing Computations Research Institute, \\
	Florida State University, Tallahassee, FL 32306-4130, USA
}
\author{Ron~Workman}
\address{
        Department of Physics
        Virginia Tech, Blacksburg, VA 24061, USA
}
\date{\today}
\maketitle

\begin{flushright}
FSU-SCRI-98-70
\end{flushright}
\def\D{\Delta}
\begin{abstract}
Motivated by recent remarks on the $\Delta^+$ mass and comparisons
between the quark model and relations based on large-$N_c$ with
perturbative flavor breaking, two sets of $\Delta$ masses consistent
with these constraints are constructed. These two sets, based either
on an experimentally determined mass splitting or a quark model of
isospin symmetry breaking, are shown to be inconsistent.
The model dependence of this inconsistency is examined,
and suggestions for improved experiments are made. An explicit quark
model calculation and mass relations based on the large-$N_c$ limit
with perturbative flavor breaking are compared. The expected level
of accuracy of such relations is realized in the quark model, except
for mass relations spanning more than one SU(6) representation.
It is shown that the $\D^0$ and $\D^{++}$ pole masses
and $\D^0-\D^+ =(\D^- -\D^{++})/3\simeq 1.5$
MeV are more consistent with model expectations
than the analogous Breit-Wigner masses and their splittings.
\end{abstract}
\def\slash#1{#1 \hskip -0.5em / }
\def\rmb#1{{\bf #1}}
\def\lpmb#1{\mbox{\boldmath $#1$}}
\def\nn{\nonumber}
\def\>{\rangle}
\def\<{\langle}
\def\D{\Delta}
\newcommand{\Eqs}[1]{Eqs.~(\protect\ref{#1})}
\newcommand{\Eq}[1]{Eq.~(\protect\ref{#1})}
\newcommand{\Fig}[1]{Fig.~\protect\ref{#1}}
\newcommand{\Figs}[1]{Figs.~\protect\ref{#1}}
\newcommand{\Sec}[1]{Sec.~\protect\ref{#1}}
\newcommand{\Ref}[1]{Ref.~\protect\cite{#1}}
\newcommand{\Refs}[1]{Refs.~\protect\cite{#1}}
\newcommand{\Tab}[1]{Table~\protect\ref{#1}}
\renewcommand{\-}{\!-\!}
\renewcommand{\+}{\!+\!}
\newcommand{\sfrac}[2]{\mbox{$\textstyle \frac{#1}{#2}$}}
%
\sloppy
\section{Introduction}
\label{Sec:Intro}

The standard $\Delta$ masses\cite{PDG} have been used in a number of
comparisons with predictions based on large-N$_c$ with perturbative
flavor breaking\cite{JL,Jenkins} and the quark model~\cite{REC,Rosner}.
The agreement generally has been poor. While
the $\Delta(1232)$ resonance has been extensively studied in both
strong and electromagnetic reactions, only the $\Delta^0$ and
$\Delta^{++}$ masses have precise values, and the $\Delta^-$
mass has never been determined.  Values for the $\Delta^0$ and
$\Delta^{++}$ masses come mainly from analyses of elastic pion-nucleon
scattering\cite{Koch,Abaev,Pest}, and the $\Delta^+$ mass has been
extracted from analyses of pion photoproduction data\cite{PDG,Miro}.

In this paper, we first note that the agreement with theory is much
improved when the $\Delta^+$ mass of Ref.\cite{Miro} is removed. The
justification for doing so has recently been clarified\cite{RW}.
Having done this, we require a pair of additional constraints to
determine the full set of $\D$ isobar masses.
The first constraint is the most reliable relation based
on large-N$_c$ and perturbative flavor breaking given by Jenkins and
Lebed~\cite{JL}, and involves only $\D$ masses. We will consider the
different sets of $\D$ isobar masses which arise from the choice of
a second constraint. One possibility is to use a linear combination of $\D$
masses determined from an analysis of elastic $\pi^{\pm}$ scattering
from the deuteron. A value for this linear combination
\begin{equation}
D \; = \; \Delta^- - \Delta^{++} + {1\over 3} \left(
         \Delta^0 - \Delta^+ \right) ,
\end{equation}
(a particle's name is used for its mass here and in what follows) has
been extracted by Pedroni {\it et al.}~\cite{Pedroni}.

Another possibility is to use a
theoretically reliable relation between the $\D$ and $\Sigma^*$ masses
from Ref.~\cite{JL}, together with a quark model estimate of the {\it
difference} between the $\Sigma^*$ and $\Sigma$ mass splittings.
As justification for this latter approach, in Sections~\ref{Sec:Masses}
and~\ref{Sec:N_c} we carefully examine the predictions of our quark
model for isospin splittings and compare with relations based on
large-N$_c$ and perturbative flavor breaking, and with the
`experimental' masses, to see where they differ. Comparisons of a
similar nature have recently been completed by
Rosner~\cite{Rosner}. The present study extends this work through the use
of a dynamical quark model which allows for SU(6) symmetry breaking in
the baryon wavefunctions, and also for non-spectator
effects~\cite{REC}, where the interactions of a pair of quarks with a
given flavor and total spin are allowed to depend on the flavor and
spin of the remaining quark. Our study also differs numerically
through the use of different experimental input.

We will show that these two different approaches give quite different
results for the $\D$ masses, which implies an inconsistency between
the measurements of the $\D^0$ and $\D^{++}$ Breit-Wigner masses, the
extracted value of $D$, and our prediction based on large $N_c$ and
the quark model.
In Section IV, we will show that the $\D^0-\D^{++}$ mass splitting
based on pole mass values is more consistent with quark model and
large $N_c$ expectations.

\section{$\D$ masses}
\label{Sec:Masses}

In the work of Jenkins and Lebed~\cite{JL}, relations between the
masses of octet and decuplet baryons are estimated at various orders
of a perturbative expansion in flavor breaking, and in powers of
$1/N_c$. The first constraint that we will use to determine a set of
$\D$ masses is
\begin{equation}
\D_3 \equiv \D^{++} - 3\D^+ + 3\D^0 - \D^- = 0 .
\label{D_3}
\end{equation}
This relation is predicted~\cite{JL} to have an accuracy~\cite{accur} of
order $\epsilon^{\prime\prime}\epsilon^\prime/N_c^3$, where
$\epsilon^\prime$ is an isospin-symmetry violating parameter for the
strong interaction mass splittings, and
$\epsilon^{\prime\prime}\simeq\epsilon^\prime$ is an isospin symmetry
breaking parameter for electromagnetic mass splittings. With the
parameters of Ref.~\cite{JL}, this means that $\D_3$ is expected to
be of order 10$^{-3}$ MeV or smaller. We will show below that our
quark model, which breaks flavor and isospin symmetry explicitly,
satisfies Eq.~(\ref{D_3}) to a similar degree of accuracy.

A set of $\D$ masses can be constructed with minimal theoretical input
by using the value of $D$ extracted by Pedroni {\it et
al.}~\cite{Pedroni} and the accurate relation Eq.~(\ref{D_3}),
\begin{eqnarray}
\Delta^0 - \Delta^+ & = & {{3 D}\over {10}} = 1.38\pm 0.06\
 {\rm MeV} ,\nn \\
\Delta^- - \Delta^{++} & = & {{9 D}\over {10} } = 4.14\pm 0.18\
 {\rm MeV} .
\label{DdiffD}
\end{eqnarray}
If these relations are combined with Breit-Wigner
masses~\cite{PDG,Koch} for the $\Delta^0$ ($1233.6\pm 0.5$ MeV) and
$\Delta^{++}$ ($1230.9\pm 0.3$ MeV), we have
\begin{eqnarray}
\Delta^+  &=& 1232.2 \pm 0.5\ {\rm MeV} , \nn \\
\Delta^-  &=& 1235.0 \pm 0.35\ {\rm MeV} .
\label{DD}
\end{eqnarray}
Although the $\Delta^+$ mass is poorly known, we note that the current
range of values, given in the Review of Particle Properties~\cite{PDG}
($1231.5\pm 0.3$ MeV, {\it excluding} the value of Ref.~\cite{Miro}),
is consistent with this value. This exercise has also been
performed by Lebed~\cite{Lebed}, with slightly different
results~\cite{Lebed2}. However, we will show in what follows that this
prescription leads to a set of masses which is in conflict with
a result based on a combination of relations derived from the large
$N_c$ limit with perturbative flavor breaking and the quark model.

\begin{figure}[t]
\centering{\ \mbox{\  \epsfig{figure=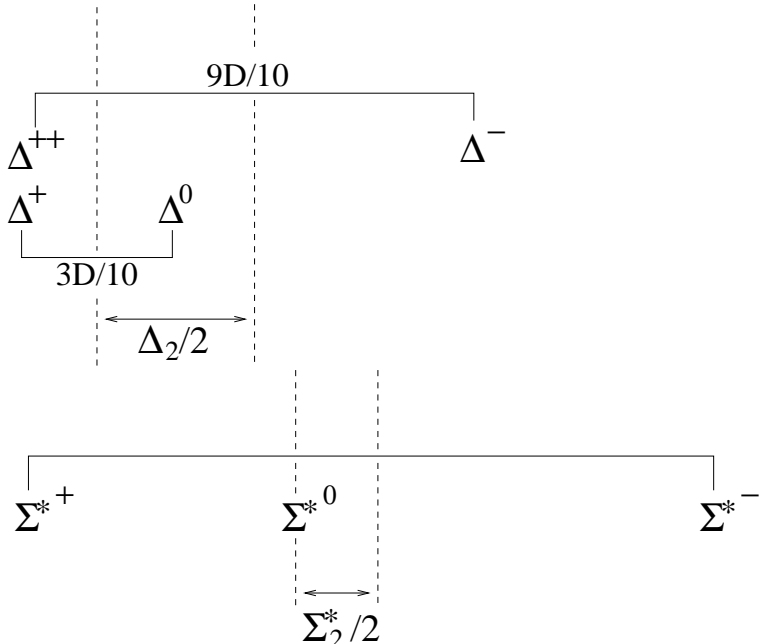,angle=-90,width=10cm}}}
\caption{$\D$ and $\Sigma^*$ mass splittings, incorporating the
relations $\D_3=0$ and $\Sigma_2^*=\D_2/2$.}
\vspace{1.0ex}
\label{Fig:splits}
\end{figure}

A relation given by Jenkins and Lebed~\cite{JL}
\begin{equation}
\D_2=2\ \Sigma^*_2 ,
\label{D2S*2}
\end{equation}
between the quantities
\begin{equation}
\Delta_2\equiv(\D^{++}+\D^-)-(\D^+ +\D^0) ,
\label{D_2}
\end{equation}
and
\begin{equation}
\Sigma^*_2\equiv\Sigma^{*+} +\Sigma^{*-} -2\Sigma^{*0} ,
\end{equation}
[with $\Sigma^*\equiv\Sigma(1385)$] can also be used to constrain the
$\D$ masses. This relation is expected to be accurate to
$\epsilon^{\prime\prime}\epsilon/N_c^3\simeq 3\times 10^{-5}$, where
$\epsilon$ is an SU(3)$_f$ symmetry violating parameter (with
$\epsilon \gg \epsilon^\prime\simeq \epsilon^{\prime\prime}$), so
that~\cite{JL} corrections to this relation should be of order 0.15
MeV. Figure~\ref{Fig:splits} illustrates the pattern of $\D$ and
$\Sigma^*$ splittings which results from imposing Eqs.~(\ref{D_3})
and~(\ref{D2S*2}).

Since the current value $\Sigma^*_2=2.6\pm 2.1$ MeV extracted from
data~\cite{PDG} is quite uncertain, our approach is to estimate
$\Sigma^*_2$ by using the value $\Sigma_2\equiv\Sigma^+ +\Sigma^-
-2\Sigma^0=1.71\pm 0.18$ MeV, also extracted from data~\cite{PDG}, and a
dynamical quark model prediction of the difference $\Sigma^*_2-\Sigma_2$.

The pairing models of baryon isospin splittings of Cutkosky~\cite{REC}
and Rosner~\cite{Rosner} assume the universality of splittings of a given
type within the ground state octet and decuplet baryons. This amounts
to the assumption of SU(6) flavor-spin symmetry in the wavefunctions,
although the interactions must be allowed to depend on the light-strange
quark mass difference and so break SU(3)$_f$. This is described in
Ref.~\cite{REC} as a spectator approximation, so that the strong and
electromagnetic interactions between a given pair of quarks do not
depend on the flavor or spin of the remaining quark in the
baryon. Models of this kind give $\Sigma^*_2=\Sigma_2$. Dynamical
models such as that used here~\cite{SC87} and by Isgur~\cite{NI} allow
for breaking of SU(6) symmetry in the wavefunctions and so allow for
and calculate non-spectator effects. However, these models neglect
potentially important additional effects due to electromagnetic box
and penguin graphs, as shown by Stephenson, Maltman, and
Goldman~\cite{SMG}, and these may also contribute to non-spectator
effects.

It is pointed out in Ref.~\cite{REC} that isospin splittings in the
hyperons are much larger than for the $N$ and $\D$ baryons due to
cancellations between pair terms in the latter which are not present
in the former. This suggests that the $N$ and $\D$ splittings may be
sensitive to non-spectator effects, which may not necessarily show the
same cancellations. It is therefore important to include these effects
when examining the $\D$, $\Sigma$, and $\Sigma^*$ mass splittings in a
model (such as ours) constrained to fit the $n-p$ and other
isospin splittings. Note that certain of the mass relations based on
large-$N_c$ and perturbative flavor breaking mentioned here are either
satisfied by construction or by virtue of the assumption of SU(6)
symmetry in the wavefunctions of the model of Refs.~\cite{REC}
and~\cite{Rosner}, and that our model allows for explicit breaking of
these relations.

As an example of such an effect, any model which consistently treats
the hyperfine contact interaction and the isospin splittings will
predict that the $\D^0-\D^+$ splitting is slightly larger than $n-p$,
because the effect of $m_d - m_u>0$ on the quark kinetic energy is
less diluted by relativistic effects in the $\D$. Explicitly,
$(m_d^2+p^2)^{1/2} - (m_u^2+p^2)^{1/2}$ is smaller than $m_d - m_u$
for a finite quark momentum $p$. The nucleon has a net attractive
contact interaction which gives the quarks a larger mean momentum,
whereas the $\D$ has a repulsive contact interaction. Slight
differences in the magnitude of the electrostatic and magnetic
interactions introduce almost no difference between these two
splittings. Although they are somewhat reduced in magnitude in the
larger $\D$ state with slower moving quarks, these two terms come in
with opposite sign and so the differences largely
cancel. The relation $\D^0-\D^+=n-p$ from Jenkins and
Lebed~\cite{JL} and Rosner~\cite{Rosner} has an expected accuracy of
$\epsilon^\prime/N_c^2$, which is significantly lower order than, say,
Eq.~(\ref{D2S*2}).

The quark model predictions given here are made within a model similar
to that of Ref.~\cite{SC87}, with some important differences noted
below. The strong contact interaction used in the baryon spectrum
calculation of Ref.~\cite{CI} was convoluted with a Gaussian smearing
function with the form ${\rm exp}(-\sigma_{ij}^2r_{ij}^2)$, where
$r_{ij}$ is the separation of quarks $i$ and $j$, and the smearing
parameter $\sigma_{ij}$ was 1.83 GeV for a light-quark pair (the
smearing parameter is taken to depend on the quark mass, see
Ref.~\cite{CI} for details). This can be interpreted as a strong form
factor for the light constituent quarks, and this smearing parameter
implies a relatively small strong size for the constituent quark. On
the other hand, relativistic calculations of the electromagnetic form
factors of the nucleon carried out with light-cone techniques require
a substantially larger electromagnetic size for the constituent quark
in order to fit the nucleon form factors using the resulting
wavefunctions~\cite{CPSS,CK,CRob}. The magnetic component of the
electromagnetic interaction {\it between} quarks, which is one source
of isospin-violating mass splittings, was smeared in Ref.~\cite{SC87},
also with a substantially smaller smearing parameter $\gamma_{\rm em}$
than that used for the strong contact interaction.

This implies that smaller smearing parameters should be used in the
strong contact interaction, coupled with a larger strong coupling
$\alpha_s(Q^2=0)$ to preserve the size of the contact splittings. This
reduces the level of high-momentum components in the nucleon, and
therefore reduces the electromagnetic size of the quarks required to
fit the nucleon moments, bringing the strong and electromagnetic
constituent quark sizes into rough agreement.

Wavefunctions have been generated for the ground state octet and
decuplet baryons with a strong contact interaction which is smeared
with $\sigma_{ij}=0.9$ GeV for light quark pairs [with similar
reductions for the $s-(u,d)$ and $s-s$ quark pairs], and with an
increased $\alpha_s(0)$, which result in a fit to the ground state
baryon and the entire light-quark baryon spectra of similar quality to
that of Ref.~\cite{CI}. The resulting wavefunctions for the nucleons
have been shown to give an adequate fit to the nucleon elastic form
factors within a light-cone model~\cite{CK,CRob}. The parameters of the
isospin splitting model of Ref.~\cite{SC87} have been readjusted to
fit the measured splittings, yielding $\delta m=m_d-m_u=3.6$ MeV and
$\gamma_{\rm em}=1.0$ GeV, with an unchanged magnetic relativistic
suppression factor $\epsilon_{\rm magn}=-0.297$. The results for
light-quark baryons are $n-p=1.3$ MeV, $D\equiv\D^-
-\D^{++}+(\D^0-\D^+)/3=4.9$ MeV, and $\D_2=3.5$ MeV. Our results confirm
within a dynamical model (constrained by the baryon spectrum, nucleon
form factors, and the measured isospin splittings) the expected
accuracy of the best of the relations based on large-$N_c$ and
perturbative flavor breaking of Ref.~\cite{JL}; we find $\D_3=0.002$
MeV and $\D_2-2\ \Sigma^*_2=-0.082$ MeV for this fit. Our quark model
explicitly breaks SU(6) symmetry, which allows a slight difference
$\Sigma_2^*-\Sigma_2=0.074$ MeV, with $\Sigma_2=1.70$ MeV, consistent with
the measured value of $1.71\pm 0.18$ MeV. These results suggest that
it should be a good approximation to constrain $\D_2$ using Eq.~(2.4)
and our quark model prediction that $\Sigma_2^*$ should be only
slightly larger than $\Sigma_2$. As a result, we will adopt the value
\begin{equation}
\D_2 = 2(1.71\pm 0.18+0.074)\ {\rm MeV.}=3.57\pm 0.36\ {\rm MeV.}
\label{ourD_2}
\end{equation}

This value of $\D_2$ is quite different from the value implied by our
first set of masses, which are based on the Breit-Wigner $\D^0$ and
$\D^{++}$ masses, $\D_3=0$ and the extracted value of $D$. To
illustrate this point, we eliminate either $\D^-$ or $\D^+$ from
Eqs.~(\ref{D_3}) and~(\ref{D_2}) to obtain
\begin{eqnarray}
\D^+&=&{\D^0 +\D^{++}\over 2}-{\D_2\over 4} , \nn\\
\D^-&=&{3\D^0 -\D^{++}\over 2}+{3\D_2\over 4},
\end{eqnarray}
which give the expressions
\begin{eqnarray}
\D^0-\D^+&=&{\D^0 -\D^{++}\over 2}+{\D_2\over 4} , \nn\\
\D^- -\D^{++}&=&3\left(\D^0-\D^+\right),
\label{DdiffD2S2}
\end{eqnarray}
where the last relation follows trivially from $\D_3=0$. Combining the
value $\D^0-\D^{++}=2.7\pm 0.6$ MeV, which results from the Breit-Wigner
masses, and Eq.~(\ref{DdiffD}) for $\D^0-\D^{+}$ which is based on
the Pedroni {\it et al.} value of $D$, we see that
Eqs.~(\ref{DdiffD2S2}) require $\D_2\simeq 0$ and so $\Sigma^*_2\simeq
0$, in conflict with Eq.~(\ref{ourD_2}).

Equivalently, inserting our value for $\D_2$ from Eq.~(\ref{ourD_2})
and the Breit-Wigner masses for $\D^0$ and $\D^{++}$ into
Eq.~(\ref{DdiffD2S2}) we find
\begin{equation}
\D^0-\D^+=2.2\pm 0.3\ {\rm MeV} .
\end{equation}
Comparing to Eqs.~(\ref{DdiffD}) we see that the effect of this
approach has been to adopt a value
$D=10(\D^0-\D^+)/3\simeq 7.5\pm 1$ MeV which
is significantly larger than that extracted by Pedroni {\it et
al.}~\cite{Pedroni}. A value of $D$ this large is also disfavored in
our quark model.

This suggests that there is an inconsistency between the Breit Wigner
values for the $\D^0$ and $\D^{++}$ masses, the value $D=4.6\pm 0.2$
MeV, and the analysis combining Eq.~(\ref{D2S*2}) with our quark model
result. Note that this argument is based on the difficulty of
accommodating substantially unequal values of $\Sigma^*_2$ and
$\Sigma_2$ in the quark model. In our quark model both $\Sigma_2$ and
$\Sigma^*_2$ have negligible contributions from the dependence of the
kinetic energy and strong interactions on the $m_d-m_u$ mass
difference. Their values result from a cancellation between a positive
Coulomb term ($\simeq 3$ MeV) and a negative magnetic term $\simeq -1$
MeV). The Coulomb and magnetic terms are slightly larger in the
spatially smaller (from the net negative contact interaction) ground
state $\Sigma$, which accounts for the slight difference between
$\Sigma^*_2$ and $\Sigma_2$. In Cutkosky's pairing model~\cite{REC},
the $m_d-m_u$ terms are exactly zero, the same partial cancellation
between electric and magnetic terms occurs, but by fiat
$\Sigma^*_2=\Sigma_2$. A value for $\Sigma^*_2$ close to zero while
$\Sigma_2$ is close to the value extracted from experiment is,
therefore, inconsistent with such quark models.  We will return to
this point in Section~\ref{Sec:BW}.

\section{Accuracy of mass relations in the quark model}
\label{Sec:N_c}

As our analysis of the $\D$ isobar masses depends crucially on the
relations in Eqs.~(\ref{D_3}) and~(\ref{D2S*2}), it is of some
interest to test the predicted accuracy of these and other relations
based~\cite{JL} on large-$N_c$ and perturbative flavor breaking with a
dynamic quark model which includes SU(3)$_f$ breaking effects, as well
as SU(6) symmetry breaking and effects higher order in the
isospin-symmetry violating quantities such as $\delta
m/m\equiv 2(m_d-m_u)/(m_u+m_d)$. Certain of these relations cannot be
compared to experiment due to large experimental uncertainties,
particularly in the splittings of the $\D$ states. Our quark model can
provide estimates for the level of accuracy of such relations.

We have already seen above that the most highly suppressed $I=2$ and
$I=3$ operators from Ref.~\cite{JL} yield relations ($\D_2=2\
\Sigma^*_2$ and $\D_3=0$ respectively) with predicted accuracies which
are realized in our model. There are also several $I=1$ mass
relations. One is $\D_1-10\ \Sigma^*_1+10\ \Xi^*_1=0$, with
$\D_1\equiv 3(\D^{++}-\D^-)+\D^+-\D^0$,\ $\Sigma_1\equiv\Sigma^+
-\Sigma^-$, and $\Xi^*_1\equiv \Xi^{*0}-\Xi^{*-}$. In our quark model
we have $\D_1-10\Sigma^*_1+10\Xi^*_1=-0.20$ MeV, which corresponds to
an accuracy of $6\times 10^{-6}$, which compares favorably with the
predicted accuracy of this relation from Ref.~\cite{JL} of
$\epsilon^\prime \epsilon^2/N_c^3\simeq 10^{-5}$. Similarly, the
Coleman-Glashow relation $N_1-\Sigma_1+\Xi_1=0$ is satisfied by our
quark model to within 0.03 MeV, which corresponds to an accuracy of
$8\times 10^{-6}$, and is predicted to be accurate~\cite{JL} to
$\epsilon^\prime \epsilon/N_c^2\simeq 10^{-4}$ in the large-$N_c$ and
SU(3)$_f$ limit.

Two additional $I=1$ relations from Ref.~\cite{JL} with an expected
accuracy of $\epsilon^\prime \epsilon/N_c^2$ are $\D_1-3\ \Sigma_1^*
-4\ \Xi^*_1=0$ and $\Sigma^*_1-2\ \Xi^*_1=0$, and are both rather
poorly satisfied by our dynamical model, which has $\D_1-3\
\Sigma_1^*-4\ \Xi^*_1=6.1$ MeV, which corresponds to an accuracy of
$4\times 10^{-4}$, and $\Sigma^*_1-2\ \Xi^*_1=-0.91$ MeV which
corresponds to an accuracy of $2\times 10^{-4}$. A similar lack of
agreement is obtained for a wide range of parameters. Both of these
relations are derived~\cite{JL} using a mass relation which does not
correspond to a single SU(6) representation. This suggests that some
mass relations which span more than one SU(6) representation, and
those derived from them, may not be consistent with our dynamical
model.

\section{Breit-Wigner versus pole masses}
\label{Sec:BW}

It is
interesting to note that our quark model fit {\it without} imposition
of constraints from the $\D$ masses gives a value $D=4.9$ MeV, close
to the Pedroni {\it et al.}  value $D=4.6\pm 0.2$ MeV. This is also true of
the fit of Cutkosky~\cite{REC}. If instead of using the Breit-Wigner
masses to determine $\D^0-\D^{++}$ we use the pole
masses~\cite{PDG,Pest,RW,Vasan}, we find a smaller splitting
\begin{equation}
\D^0-\D^{++}\simeq 1\ {\rm MeV} .
\label{poleD0D++}
\end{equation}
A similar result for this splitting was found by Cutkosky\cite{REC} in
a fit to the octet and decuplet baryons which excluded the $\D$ masses.
Evaluating
Eq.~(\ref{DdiffD2S2}) with the pole mass difference and our value of
$\D_2$ gives
\begin{eqnarray}
\D^0-\D^+&\simeq &1.5\ {\rm MeV} ,\nn\\
\D^- -\D^{++}&\simeq &4.5\ {\rm MeV} ,
\end{eqnarray}
which is at least compatible with our quark model expectation that $\D^0-\D^+$
should be slightly larger than $n-p=1.3$ MeV, though the uncertainty
associated with pole mass splittings cannot support any more quantitative
conclusions. This naturally leads one
to consider whether pole or Breit-Wigner masses should be used in mass
relations. As has been pointed out by H\"ohler\cite{Hoehler}, the pole
position (and not the Breit-Wigner mass) is a quantity which can be
most rigorously associated with a resonance.

One obvious way to address this question is to use the best $I=0$ mass
relation from Ref.~\cite{JL} with an expected accuracy of
$\epsilon^3/N_c^3$,
\begin{equation}
\Delta_0 = 3\left( \Sigma_0^* - \Xi_0^*\right) + \Omega_0,
\label{D_0}
\end{equation}
where $B_0$ is the average of the isobar masses of baryon $B$. This
leads, unfortunately, to a central value between the $\Delta$
Breit-Wigner and pole (1210 MeV) masses. However, Dillon has
noted\cite{Dillon} that the consistency of Eq.~(\ref{D_0}) with the
$\D$ pole masses is improved if pole values\cite{Lich} are used
consistently for each particle. This distinction has no effect on the
$\Omega$ mass, but does shift the $\Sigma^*_0$ and $\Xi^*_0$ terms
slightly.  Here again, improved values for the $\Sigma^*$ (and
$\Xi^*$) Breit-Wigner and pole masses would lead to a corresponding
improvement in our understanding of the $\Delta$.

While the $\Delta$ ``mass'' has been variously quoted near 1210 MeV,
1232 MeV, and even 1241 MeV\cite{Dillon,CCD,Cheng}, the differences are
mainly due to model dependence in the separation of resonance and
background contributions. The pole position remains stable near
1210$-i$50 MeV in all of these works. As an exercise, we have repeated
our quark model calculation of the isospin splittings but with the
strong contact interaction altered to fit $\D-N$ evaluated with the
$\D$ pole mass, leaving all other details of the model unchanged. The
resulting values of $n-p$, $D$, and $\Delta_2$ were largely unchanged
at $1.3$ MeV, $4.8$ MeV, and $3.5$ MeV respectively, although the magnitudes
of the splittings $\Sigma_1$,\ $\Sigma^*_1$, and $\Xi_1$ were
reduced. This simple exercise suggests that it may be possible to
accommodate the average pole masses and their difference in a quark
model of this kind.

\section{Conclusions}
\label{Sec:Concl}

Two different constructions for the $\D$ isobar masses are
compared. Both are based on the well determined $\D^0$ and $\D^{++}$
masses, and a relation based on perturbative flavor breaking and
large-$N_c$, with an expected very high degree of accuracy realized in
our dynamical quark model. The additional constraint is taken to be
either a combination of $\D$ masses extracted from $\pi^{\pm}$
deuteron elastic scattering data, or a second large-$N_c$ relation in
combination with a quark model calculation. The expected high degree
of accuracy of this second large-$N_c$ relation is also realized in
our explicit quark model. If the Breit-Wigner $\D^0$ and $\D^{++}$
masses are adopted, these two different constructions are in
conflict. The quark model relation, upon which this conflict is based,
is a basic consequence of the type of model used, and does not depend
sensitively on parameter values.

The accuracy of certain relations based on perturbative flavor
breaking and large-$N_c$ is unknown because of uncertainties in the
masses extracted from the data. We have found that in most cases the
predicted accuracy is realized in our model. The
exceptions are relations which are not restricted to a single SU(6)
representation. We have also shown that the relations between masses
based on perturbative flavor breaking and large-$N_c$ and those
derived from the quark model are more consistent with the $\D$ pole
masses than the corresponding Breit-Wigner values. This suggests that
$\D^0-\D^+ =(\D^- -\D^{++})/3\simeq 1.5$ MeV and the pole
mass difference $\D^0-\D^{++}\simeq 1$ MeV are consistent with
theoretical expectations.

We support Rosner's assertion that improved $\Sigma^*$ and
$\Xi^*$ masses are vitally important. These would sharpen
comparisons between the large-$N_c$ and quark model predictions
and allow more quantitative comparisons between mass relations
using Breit-Wigner and pole masses. With one additional $\D$
mass, we could use Eq.~(2.1) to determine the remaining state,
bypassing the deuteron data. While the $\D^+$ mass determined
from pion photoproduction data is the most obvious candidate,
we should point out a potential problem. This is most obvious
if we rewrite Eq.~(2.1) in the form
\begin{equation}
\D^- \ = \ \D^{++} \ + \ 3 \left( \D^0 - \D^+ \right) ,
\end{equation}
which would be utilized, given the $\D^0$ and $\D^{++}$ masses.
Since $\D^0$ has a larger uncertainty than $\D^{++}$ and the
`expected' $\D^0 - \D^+$ splitting is only about 1.5 MeV, we
could easily have an experimental $\D^0 - \D^+$ splitting
consistent with zero. This uncertainty would then be magnified
in our estimate of the $\D^-$ mass. While a direct measurement
of the $\D^-$ mass would have the greatest impact, it is
unfortunately the least favorable experiment, involving the
extraction of $\pi^- n$ scattering from a deuteron target.
\section*{Acknowledgments}
One of us (S.C.) would like to acknowledge helpful discussions with
Prof.~N.~Isgur and Prof.~D.~Robson about the contact interaction in
baryons.  This work is supported by the U.S. Department of Energy
under Contract DE-FG02-86ER40273 and the Florida State University
Supercomputer Computations Research Institute which is partially
funded by the Department of Energy through Contract DE-FC05-85ER25000
(S.C.), and by the U.S. Department of Energy under Contract
DE-FG02-97ER41038 (R.W.).


\begin{references}
\bibitem{accur} In Refs.~\cite{JL} and~\cite{Rosner} and in this work,
   the accuracy of a mass relation is defined by first rearranging the
   relation so that it reads LHS = RHS, with only positive mass terms
   on both sides, and then accuracy $\equiv$ 2 $\vert$ LHS-RHS $\vert$/(LHS+RHS).
\end{references}
\end{document}